\documentclass[aps,preprint,nofootinbib]{revtex4-1}
\usepackage{graphicx}
\usepackage{amsmath}
\usepackage{amscd}
\usepackage{amssymb}
\usepackage{latexsym}

\newsavebox{\tallguy}
\savebox{\tallguy}{\mbox{\rule{0ex}{2.25ex}}}

\newcommand{\ket}[1]{ \usebox{\tallguy} \left | #1 \right \rangle}
\newcommand{\bra}[1]{ \left \langle #1 \right | \usebox{\tallguy}}

\newcommand{\proj}[1]{\ket{#1} \! \bra{#1}}

\newcommand{\ave}[1]{\left \langle #1 \right \rangle}
\newcommand{\hilbert}{\mathcal{H}}

\newcommand{\oper}[1]{\boldsymbol{#1}}

\newcommand{\sys}[1]{^{\mbox{\tiny (#1)}}}
\newcommand{\subtext}[1]{_{\mbox{\tiny #1}}}

\newcommand{\stateset}{\mathcal{S}}
\newcommand{\kinset}{\mathcal{K}}
\newcommand{\statementset}{\mathcal{I}}

\newcommand{\symgroup}{\mathcal{U}}

\begin{document}

\title{Interpretation of quantum theory:  the quantum ``grue-bleen'' problem}
\date{\today}

\author{Benjamin Schumacher}
\affiliation{Department of Physics, Kenyon College, Gambier, OH 43022}
\email[Corresponding author: ]{schumacherb@kenyon.edu}
\author{Michael D. Westmoreland}
\affiliation{Department of Mathematics and Computer Science, Denison University, Granville, OH 43023}

\begin{abstract}
	We present a critique of the many-world interpretation of quantum
	mechanics, based on different ``pictures'' that describe the time evolution of
	an isolated quantum system.  Without an externally imposed frame to restrict 
	these possible pictures, the theory cannot yield non-trivial interpretational 
	statements.  This is  analogous to Goodman's famous ``grue-bleen'' 
	problem of language and induction.  Using a general framework applicable
	to many kinds of dynamical theories, we try to identify the kind of
	additional structure (if any) required for the meaningful interpretation of a theory.
	We find that the ``grue-bleen'' problem is not restricted to quantum
	mechanics, but also affects other theories 
	including classical Hamiltonian mechanics.  For all such theories,
	absent external frame information, an isolated system has no interpretation.
\end{abstract}

\maketitle

\section{Introduction}

\subsection{The many-worlds interpretation}

Any critique of the many-worlds interpretation of quantum mechanics
ought to begin by praising it.  In the simplest form of the interpretation, 
such as that presented by Everett in 1957 
\cite{everett1957}, 
the universe is regarded as a closed quantum
system.  Its state vector (Everett's ``universal wave function'') evolves 
unitarily according to an internal Hamiltonian.  Measurements
and  the emergence of classical phenomena are described entirely
by this evolution.  ``Observables'' are simply dynamical variables
described by operators.  No separate ``measurement process'' or ``wave
function collapse'' ideas are invoked.

Thus, consider a laboratory measurement of $S_{z}$ on 
a spin-1/2 particle.  This is nothing more than an interaction 
among the particle, the lab apparatus,
and the conscious observer, all of which are subsystems of
the overall quantum universe.  Initially, the particle is in the
state $\ket{\psi_{0}} = \alpha \ket{\uparrow} + \beta \ket{\downarrow}$.
The apparatus and the observer are in initial states $\ket{0}$ and 
$\ket{\mbox{``ready''}}$ respectively.  Now the particle and the
apparatus interact and become correlated:
\begin{equation}
	\ket{\psi_{0}} \otimes \ket{0} \otimes \ket{\mbox{``ready''}} \longrightarrow
	\Big ( \alpha \ket{\uparrow} \otimes \ket{+\mbox{$\frac{\hbar}{2}$}}
		+ \beta \ket{\downarrow} \otimes \ket{-\mbox{$\frac{\hbar}{2}$}} \Big )
		 \otimes \ket{\mbox{``ready''}},
\end{equation}
where $\ket{+\mbox{$\frac{\hbar}{2}$}}$ and $\ket{- \mbox{$\frac{\hbar}{2}$}}$ are
apparatus states representing the two possible measurement results.
The observer next interacts with the apparatus by reading its output, 
leading to a final state
\begin{equation}  \label{eq-twobranches}
	\alpha \ket{\uparrow} \otimes \ket{+\mbox{$\frac{\hbar}{2}$}}
		\otimes \ket{\mbox{``up''}} +
		 \beta \ket{\downarrow} \otimes \ket{-\mbox{$\frac{\hbar}{2}$}}
		\otimes \ket{\mbox{``down''}} .
\end{equation}
The memory record of the observer (``up'' or ``down'')
has become correlated to both the original spin and the reading on
the apparatus.  The two components of the superposition in
Equation~\ref{eq-twobranches} are called ``branches''
or ``worlds''.  Since
all subsequent evolution of the system is linear, the branches
effectively evolve independently.  The observer can condition
predictions of the future behavior of the particle on his own memory
record---for example, if his memory reads ``spin up'' then he may 
regard the state of the spin as $\ket{\uparrow}$.  No collapse has
occurred; both measurement outcomes are still present in the
overall state.  But conditioning on a particular memory record
yields a {\em relative state} of the particle that corresponds to
that record.  In the same way, if other observers read 
the apparatus or perform independent measurements 
of the same observable, all observers will
find that their memory records are consistent.

Here is another way to look at this process.  Consider the dynamical
variable $\oper{C}$ on the spin-observer subsystem given by:
\begin{equation}
	\oper{C} = \proj{\uparrow} \otimes \proj{\mbox{``up''}} + 
		\proj{\downarrow} \otimes \proj{\mbox{``down''}} .
\end{equation}
This variable is a projection onto the subspace
of system states in which the spin state and the observer
memory state agree.  At the start of the measurement process,
the ``expectation'' $\ave{C} = \bra{\Psi} \oper{C} \ket{\Psi} = 0$, 
but at the end $\ave{C} = 1$.
The evolution of $\ave{C}$ tells us that a correlation has emerged
between the spin and the memory record.  Note that this does
not depend on a probabilistic interpretation of the expectation
$\ave{C}$.  The expectation $\ave{C}$ simply indicates the 
relationship between the system state and eigenstates of $\oper{C}$
that are either uncorrelated ($\ave{C} = 0$) 
or correlated ($\ave{C} = 1$).

There are many things to like about the many-worlds account.  
It entails no processes
other than the usual dynamical evolution according to the Schr\"{o}dinger 
equation.  It explains at least some characteristics of a measurement,
such as the repeatability and consistency of the observers' records.
It focuses attention on the actual physical interactions involved in the
measurement process.  Some details may be tricky, such as the 
identification of $|\alpha|^{2}$ and $|\beta|^{2}$ as observed outcome 
probabilities in repeated measurements \cite{bornrule}.  
Nevertheless, the many-worlds
idea has proven to be very fruitful, for example, in motivating the
analysis of decoherence processes \cite{zurek}
and their role in the emergence 
of quasi-classical behavior in quantum systems \cite{quasi}.

The essential idea of the many-worlds program was
formulated by Bryce DeWitt \cite{dewitt} in the following maxim:
\begin{quotation}
	The mathematical formalism of the quantum theory
	is capable of yielding its own interpretation.
\end{quotation}
DeWitt called this the ``EWG metatheorem'', after Everett
and two other early exponents of the interpretation, John 
Wheeler \cite{wheeler} and Neill Graham \cite{graham}.  
DeWitt's claim is that the only
necessary foundations for sensible interpretational statements
about quantum theory are already present in the mathematics
of the Hilbert space of states and the time evolution of the
global system.  Nothing outside of the system and its 
unitary evolution is required.

\subsection{Two universes, two pictures} \label{subsec:twouniverses}

Consider a closed quantum ``universe'', which we will call Q.
System Q is composite with many subsystems.  Its time evolution
is unitary, so that the state at any give time is
\begin{equation}
	\ket{\Psi(t)} = \oper{U}(t) \ket{\Psi_{0}} 
\end{equation}
for evolution operator $\oper{U}(t)$ and initial state
$\ket{\Psi_{0}}$.  For convenience, we will refer to this as the
``actual'' time evolution of the system.

To make our mathematical discussion straightforward, 
we imagine that Q is bounded in space, so that its Hilbert space
$\hilbert\sys{Q}$ has a discrete countable basis set.
(The Hamiltonian eigenbasis would be an example of such.)  
If we further impose an upper limit $E_{\max}$ to the allowed energy
of the system, the resulting $\hilbert\sys{Q}$ is finite-dimensional.
Note that this scarcely limits the possible complexity of Q.  The system
may still contain a multitude of subsystems with complicated
behavior.  The subsystems may exchange information and energy.  
Some of the subsystems may function as ``observers'', interacting with
their surroundings and recording data in their internal memory
states.

According to the DeWitt maxim, the initial state $\ket{\Psi_{0}}$
and time evolution operator $\oper{U}(t)$ suffice to specify
a many-worlds interpretation of what happens in Q.  One way to describe
this is to consider a large collection of dynamical variables 
$\oper{A}_{1}$, $\oper{A}_{2}$, etc.  These may 
represent particle positions, observer memory states, correlation
functions, and so on.  From the time-dependent expectations
$\ave{A_{k}}_{t}$ we identify processes such as measurements,
decoherence, and communication.\footnote{Indeed, if the set 
$\{ \oper{A}_{k} \}$ is large enough, we can completely reconstruct the
time evolution $\ket{\Psi(t)}$ from the expectations $\ave{A_{k}}_{t}$.}  
We can in principle tell what the system ``looks like'' to various 
observer subsystems inside Q.

We next introduce a different, much simpler closed system Q$'$ consisting
of three coupled harmonic oscillators.  Again the Hilbert space
$\hilbert\sys{Q$'$}$ has a discrete countable basis, and if we 
further impose an upper energy limit we can arrange for
$\dim \hilbert\sys{Q} = \dim \hilbert\sys{Q$'$}$.  The two Hilbert
spaces are therefore isomorphic, and there exists an isomorphism 
map for which the initial Q$'$ state corresponds to the initial Q state.  
This means we can effectively regard 
Q and Q$'$ as the {\em same} system with the same initial
state $\ket{\Psi_{0}}$ evolving under different
time evolutions $\oper{U}(t)$ and $\oper{V}(t)$.  
Variables $\oper{B}_{k}$ for Q$'$ are different
operators in $\hilbert\sys{Q}$, corresponding to the 
oscillator positions and momenta, etc.  With respect to the 
alternate $\oper{V}(t)$ evolution, the expectations of these
Q$'$ variables would be
\begin{equation}
	\ave{B_{k}}'_{t} =  \bra{\Psi_{0}} \oper{V}^{\dagger}(t) \oper{B}_{k} \oper{V} \ket{\Psi_{0}}.
\end{equation}
These expectations would tell us ``what happens'' in Q$'$.
(The actual evolution of $\ave{B_{k}}_{t}$ under the 
actual time evolution $\oper{U}(t)$ would, of course,
be quite different.)

Now consider a new set of variables in Q:
\begin{equation}
	\tilde{\oper{B}}_{k} = 
		\left ( \oper{U}(t) \oper{V}^{\dagger}(t) \right ) \oper{B}_{k}
		\left (\oper{V}(t) \oper{U}^{\dagger} (t) \right ) .
\end{equation}
The $\tilde{\oper{B}}_{k}$ operators are time dependent.
But consider how their expectations evolve in time under the
actual time evolution of Q.
\begin{eqnarray*}
	\ave{\tilde{B}_{k}}_{t} & = & \bra{\Psi_{0}} \oper{U}^{\dagger}
		\tilde{\oper{B}}_{k} \oper{U} \ket{\Psi_{0}}
		\nonumber \\
		& = & \bra{\Psi_{0}} \oper{U}^{\dagger} \oper{U} \oper{V}^{\dagger}
			\oper{B}_{k} \oper{V} \oper{U}^{\dagger} \oper{U} \ket{\Psi_{0}}
		\nonumber \\
		& = & \bra{\Psi_{0}} \oper{V}^{\dagger} \oper{B}_{k} \oper{V} \ket{\Psi_{0}},
\end{eqnarray*}
exactly the time dependence of $\ave{B_{k}}'_{t}$ under the 
alternate Q$'$ time evolution $\oper{V}$.  In other
words, with respect to these time-dependent variables,
the complex system Q behaves exactly like the much simpler 
system Q$'$.

There is nothing particularly strange about considering time-dependent
observables.  We have described Q and its evolution using the
{\em Schr\"{o}dinger picture} \cite{asher}, in which observables are typically 
time-independent and system states evolve in time.  But we can
also use the  equivalent (and only slightly less familiar)
{\em Heisenberg picture}, in which time dependence is shifted
to the observables.\footnote{The time-dependence of observables
in the Heisenberg picture has conceptual appeal.  After all, to measure a
particle's spin on Monday or on Tuesday would require slightly
different experimental set-ups, and so the two observables
may plausibly be represented by different operators.}
The system state is thus $\ket{\Psi_{0}}$
at all times but the observables are redefined as
\begin{equation}
	\hat{\oper{A}}_{k} (t) = \oper{U}^{\dagger}(t) \oper{A}_{k} \oper{U}(t) .
\end{equation}
Then $\ave{A_{k}}_{t} = \bra{\Psi (t)} \oper{A}_{k} \ket{\Psi (t)}
= \bra{\Psi_{0}} \hat{\oper{A}}_{k}(t) \ket{\Psi_{0}}$.  In perturbation
theory, we also frequently use an {\em interaction picture}, in which
the time evolution due to an unperturbed Hamiltonian $\oper{H}_{0}$
is shifted to the observables, while the interaction Hamiltonian
$\oper{H}\subtext{int}$ produces changes in the system state.

What we have done, therefore, is simply changed pictures.
With respect to the time-dependent variables $\tilde{\oper{B}}_{k}(t)$ 
in the {\em Q$'$ picture}, the actual time evolution of Q exactly matches
the hypothetical time evolution of Q$'$.  And of course, we can generalize
this idea.  For {\em any} closed Q$'$ with a Hilbert space of the same dimension 
as $\hilbert\sys{Q}$, and for any hypothetical Q$'$ time evolution $\oper{V}(t)$,
we can find a set of time-dependent variables
with respect to which the actual Q time-evolution looks like the alternate
Q$'$ evolution.  Complex universes can be made to look simple and 
{\em vice versa}.

\subsection{Grue and bleen}

Our argument calls to mind an idea from philosophy,
devised in 1955 by Nelson Goodman.\cite{goodman}
We begin with familiar
terms {\em blue} and {\em green} describing the colors of objects
in our surroundings.  Now we fix a time $T$ and define new terms
{\em grue} and {\em bleen} as follows.
\begin{itemize}
	\item  An object is {\em grue} if is {\em green} before $T$ and {\em blue} after.
	\item  An object is {\em bleen} if it is {\em blue} before $T$ and {\em green} after.
\end{itemize}
Goodman presented this idea to illustrate his ``new riddle of 
induction''.  If we fix $T$ to lie in the future, then all present
evidence that an object is {\em green} is also evidence that it is {\em grue}.  
Here, however, we are not principally concerned about inductive
reasoning.  It does not matter to us whether 
$T$ lies in the future or the past.

In the quantum situation, the ordinary Q-observables $\oper{A}_{k}$
correspond to the ordinary colors {\em green} and {\em blue}.  
The time-dependent Q$'$-picture observables $\tilde{\oper{B}}_{k}$ 
correspond to the new terms {\em grue} and {\em bleen}.
\begin{figure}
\begin{center}
\includegraphics[width=5.0in]{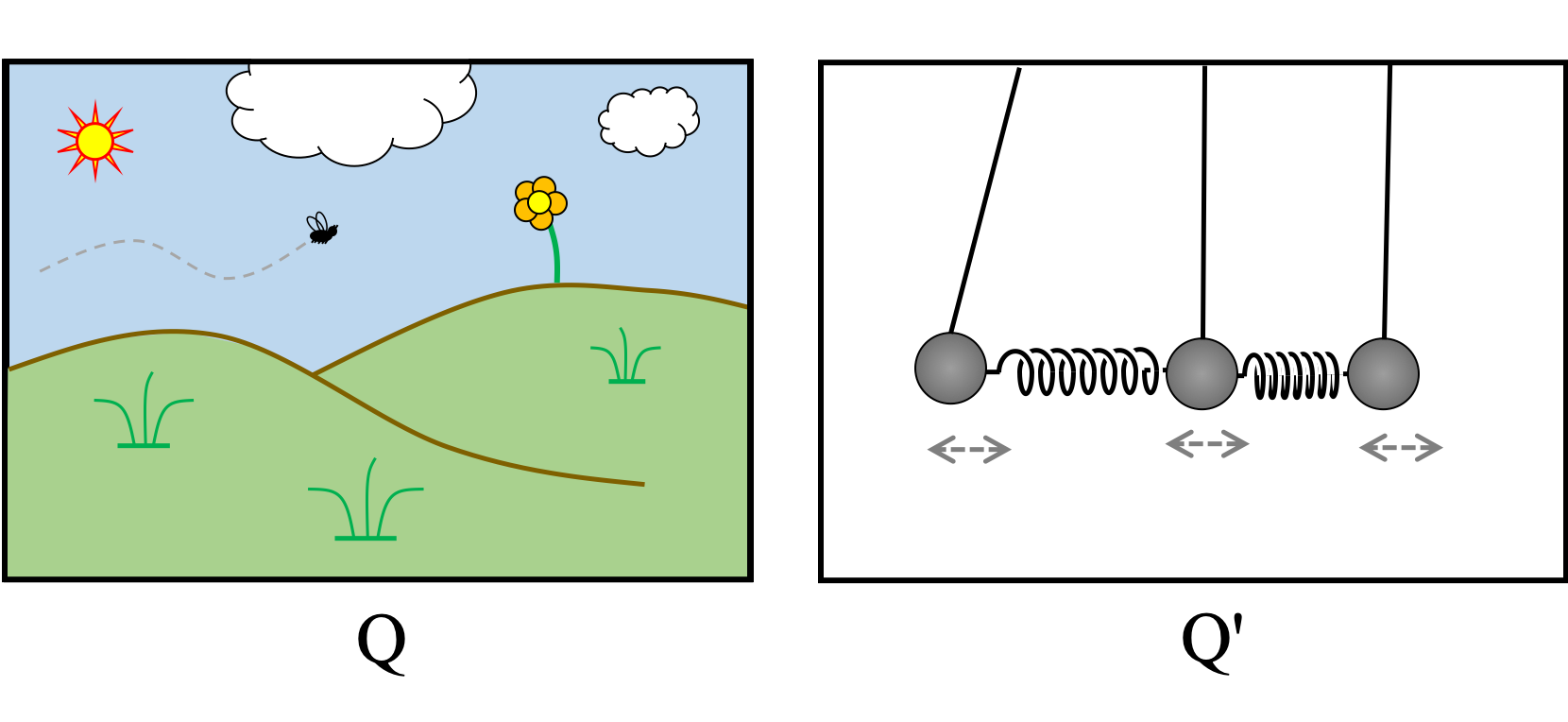}
\end{center}
\caption{Two universes.  Q is complex and contains many subsystems, including
those that may be regarded as observers (such as the bee).
Q$'$ is extremely simple.  Nevertheless, the two Hilbert spaces $\hilbert\sys{Q}$
and $\hilbert\sys{Q$'$}$ are isomorphic, so that Q and Q$'$ may be regarded as
two pictures of the {\em same} universe.}
\end{figure}

We have an intuition that the terms {\em grue} and {\em bleen} are less basic
than {\em green} and {\em blue}.  After all, the definitions of {\em grue} and 
{\em bleen} are explicitly time-dependent.  
On the other hand, suppose we start with 
{\em grue} and {\em bleen} and pose these time-dependent definitions:
\begin{itemize}
	\item  An object is {\em green} if it is {\em grue} before $T$ and {\em bleen} after.
	\item  An object is {\em blue} if is it {\em bleen} before $T$ and {\em grue} after.
\end{itemize}
Thinking only about the language, the best we can do is say that
the {\em green-blue} system and the {\em grue-bleen} system are 
time-dependent {\em relative to each other}.

In the same way, we could begin with the $\tilde{\oper{B}}_{k}$ 
description and define the Q-picture $\oper{A}_{k}$ operators 
as time-dependent combinations of them.  Each set of 
observables is time-dependent with respect to the other.

We can distinguish the two color systems by going outside mere 
language and considering the operational meaning of the terms.
We can define green and blue by a measurement of, say 
light wavelength.  To determine whether an object is green,
we can use a similar operational procedure both before and
after time $T$.  But the procedure to determine whether the object
is grue will work differently before and after $T$.  It is this
appeal to external facts that makes the green-blue distinction
more basic and elementary than the grue-bleen distinction.

What can we say about our Q and Q$'$ pictures?  We might
appeal to the physical measurement procedures required to
measure $\oper{A}_{k}$ and $\tilde{\oper{B}}_{k}$.  The procedure
for measuring $\oper{A}_{k}$ is simple and time-independent,
while that for measuring $\tilde{\oper{B}}_{k}$ is complicated 
and changes with time.  But as long as we only consider 
measurement devices and processes within our closed quantum
system, this does not suffice.  $\tilde{\oper{B}}_{k}$ devices and
processes would be simple and time-independent in the Q$'$ picture,
while $\oper{A}_{k}$ devices and processes would be wildly 
time-varying in the same picture.

This is a reference frame problem.  
In both Galilean and Einsteinian relativity, there is no natural, universal 
way to identify points in space at different times.  Space is too 
smooth and uniform; it does not have intrinsic ``landmarks''.  Hence, there is no
natural and universal way to determine whether an object is ``at
rest''.  In the same way, the Hilbert space $\hilbert\sys{Q}$ is also
too smooth and uniform to identify state vectors and operators
at different times.  From within the system, we cannot determine
whether a given collection of observables is time-dependent.

If we cannot distinguish the Q and Q' pictures from within the system,
the natural thing is to appeal to hypothetical measurement 
devices external to Q, unaffected by our change of picture.  Then
$\oper{A}_{k}$ devices are objectively simpler than $\tilde{\oper{B}}_{k}$
devices.  But this appeal to something {\em outside} of the 
closed system Q is explicitly excluded by DeWitt's maxim.
We appear to be left with an inescapable dilemma.  If we can
only consider how the state of the system evolves, then that
same history $\ket{\Psi(t)}$ can appear, with respect to different
pictures, as either the complex system Q or the simple
system Q$'$ {\em or any other quantum system with the same
Hilbert space, undergoing any unitary time evolution whatsoever}.  
We cannot identify one of these pictures as the ``correct'' one
without appealing to external measurement devices---that is,
to measurement apparatus not treated as part of the isolated
quantum system.


\subsection{What is a system?}. \label{subsec-whatisasystem}

Since the Hilbert spaces of quite different quantum systems are 
isomorphic, some additional information is required to apply quantum
theory in an unambiguous way.  This is not a novel point.  For example,
David Wallace \cite{wallace} says, ``[A]bsent additional structure, a
Hilbert-space ray is just a featureless, unstructured object, whereas
the quantum state of a complex system is very richly structured.''
Wallace regards this additional structure as part of the specification
of the quantum system in the first place.  He considers two possible
ways the provide this structure:  a specified decomposition of the
quantum system into subsystems (and thus its Hilbert space into 
quotient spaces), or a specified set of operators of fixed meaning.
In this view, the two universes Q and Q', with sets of operators 
$\{ \oper{A}_{k} \}$ and $\{ \tilde{\oper{B}}_{k} \}$, are entirely
different systems rather than different pictures of the same system.

The rest of this paper has two aims.  First, we want to pin down the
nature of the additional structure that Wallace posits.  We will
do this by considering the problem in more generality.
Section~\ref{sec-framework} presents a general framework
for describing theories that include states, time evolution, and
interpretational statements.  Such a framework naturally entails
groups of automorphisms, which we examine in Section~\ref{sec-similarities}.
Some theories, including both quantum and classical mechanics, 
require ``frame information'' to resolve ambiguities that arise
from these automorphisms.  Section~\ref{sec-examples} presents
several examples of our framework in action.

In Section~\ref{sec-taming} we turn to our second aim, which is to
use our general framework to evaluate the additional structure
required for a meaningful interpretation (of the many-worlds variety or not).
What this physical nature of this frame information?  In what ways
might the strict many-worlds program---as embodied by De Witt's 
maxim---prove inadequate?  Section~\ref{sec-remarks} includes remarks and
observations occasioned by a our line of reasoning.

\section{A general framework} \label{sec-framework}

\subsection{States and time evolution}

A {\em schema} for a theory has several parts.  We begin with a set of {\bf states}
$\stateset = \{ x, y, z, \ldots \}$.  Informally, these might be definite states or, in
the case of a non-deterministic theory, probability distributions over collections
of definite states.

To model time evolution, we introduce a sequence $(t_0, t_1, \ldots, t_{N})$
of times, where $N \geq 1$.  
Each time $t_{k}$ is associated with a state $x_k = x(t_{k}) \in \stateset$.
The whole sequence $\vec{x} = (x_0, x_1, \ldots, x_N)$ may be termed a
{\em trajectory}.  Our schema includes a set of 
{\bf kinematically possible maps} $\kinset = \{ D, E, \ldots \}$, which
are functions on the set of states:  $D: \stateset \rightarrow \stateset$ for $D \in \kinset$.
(To avoid a proliferation of parentheses, we will denote the action of $D$ on state $x$
as $Dx$ rather than $D(x)$.)
The maps in $\kinset$ describe the evolution of the state over each 
interval in our time sequence.  Thus, for the interval from $t_{k}$ to $t_{k+1}$,
\begin{equation}
	x_{k+1} = D_{k+1,k} \, x_{k} 
\end{equation}
for some $D_{k+1,k} \in \kinset$.  The sequence 
$\vec{D} = (D_{1,0}, D_{2,1}, \ldots, D_{N,N-1})$ thus describes
the time evolution over the entire sequence of time intevals.
A pair $(x_0, \vec{D})$ includes an initial state $x_0 \in \stateset$
and a sequence $\vec{D} \in \kinset^{N}$ 
of time evolution maps; such a pair is called a specific {\em instance} 
of the theory.

We can of course compose successive maps.
In the general case we do not assume that $\kinset$ is
closed under composition, so it may be that 
$D_{k+2,k} = D_{k+2,k+1} D_{k+1,k}$ is not in $\kinset$.
But in many specific cases, $\kinset$ actually
forms a group, being closed under composition and containing
both the identity map 1 and inverses for every element.
In such cases, we say that our theory is {\em reversible}.
In a reversible theory, $\kinset$ includes maps between any
pair of times $t_j$ and $t_k$, where $j,k \in \{0,\ldots,N\}$:
\begin{equation}  \label{eq-dkjdef}
	D_{k,j} = \left \{ 
		\begin{array}{ll}
		D_{k,k-1} \cdots D_{j+1,j}	& k > j \\
		1 & k = j \\
		D_{j,k}^{-1} & k < j .
		\end{array}
	\right .
\end{equation}
The algebraic structure of $\kinset$ is reflected in the way
that maps combine.  If $\kinset$ is a group, then
for any $j,k,l \in \{0,\ldots,N\}$, we have 
\begin{equation}  \label{eq-composition}
	D_{k,j} = D_{k,l} D_{l,j} .
\end{equation}
(Note that, in a reversible theory, this relation holds for any
time order of $t_{j}$, $t_{k}$ and $t_{l}$.)

If $\kinset$ is a group, it is not hard to generalize our schema
to a continuous time variable $t$.  A trajectory is a function $x(t)$
that yields a state in $\stateset$ at any time.  For any two times
$t_{1}$ and $t_{2}$, we have a map $D(t_{2},t_{1})$ such that
$x(t_{2}) = D(t_{2},t_{1}) x(t_{1})$.  These maps are related to 
one another by a composition relation analogous to 
Equation~\ref{eq-composition}:
\begin{equation}
	D(t_{2},t_{1}) = D(t_{2},t_{3}) D(t_{3},t_{1}) .
\end{equation}
Everything in the schema works pretty much the same.
For ease of exposition we will base our discussion 
on a finite sequence of discrete times $(t_{0}, \ldots , t_{N})$,
leaving the straightforward generalization to continuous
time schemata for the reader.

At the other end of the ``time complexity spectrum'',
our later examples of our framework will involve only a single
time interval from $t_0$ to $t_1$.  The set $\kinset$
may still be closed in these schemata, or even have a
group structure, but the composition of maps 
will not correspond to time evolution over
successive intervals.

\subsection{Interpretational statements}

What is an interpretation?  To give a general answer to this question is beyond
the scope of this paper.    We will merely assume that every theory
comes equipped with a collection $\statementset$ of {\bf interpretational statements}, 
which are propositions about the state and/or the map of a particular instance
of the theory.  
For example, immediately after giving the 
quantum state in Equation~\ref{eq-twobranches},
we stated, {\em The memory record of the observer (``up'' or ``down'')
has become correlated to both the original spin and the reading on
the apparatus.}  This is an interpretational statement, and its truth
is determined by the properties of the state in Equation~\ref{eq-twobranches}.
In our abstract framework, we will not be much concerned with the
{\em content} of an interpretational statement, but rather with the 
fact that it is a statement about elements of the mathematical 
formalism of our theory.  Thus, a state proposition is a statement $P(x)$ 
about a state $x \in \stateset$, and a more general type 
of proposition would be $P(x_{0},\vec{D})$, referring to an 
initial $x_{0} \in \stateset$ and a sequence of time evolution
maps $\vec{D}$.  (Notice that the more general form also 
encompasses propositions about states at any time $t_{k}$, since
we can construct the entire state trajectory $\vec{x}$ from
$x_0$ and $\vec{D}$.)
Statements of both kinds may appear in $\statementset$.
Whatever else an interpretation may include, it must surely entail 
such a set of interpretational statements; 
and if this set is empty or trivial, the interpretation is nugatory.

An interpretational statement is either true or not true.  
We say ``not true'' here rather than ``false'', because it may
be that a statement has an indeterminate value.  Consider a naive example.
For a spin-1/2 particle, our statement $P$ is ``$S_{z} = +\mbox{$\frac{\hbar}{2}$}$.''  If
the spin state is $\ket{\uparrow}$, the statement $P$ is true, inasmuch as a
measurement will surely confirm it.  If the spin state is $\ket{\downarrow}$, it is
reasonable to call $P$ false, since its negation 
(``$S_{z} \neq +\mbox{$\frac{\hbar}{2}$}$'') is true in the
same sense.  But if the spin state is $\ket{\rightarrow}$, neither $P$ nor 
its negation is true.  Thus, we simply say that $P$ is true for the state
$\ket{\uparrow}$ and not true for other states like $\ket{\downarrow}$ 
and $\ket{\rightarrow}$.

Without a more explicit ``theory of interpretation'', we cannot say
more about the structure of $\statementset$.  For example, we
do not assume that the collection $\statementset$ has any particular
algebraic closure properties.  If $P,Q \in \statementset$, we have no warrant to 
declare that $\neg P$, $P \vee Q$, or $P \wedge Q$ are part of $\statementset$.

\section{Similarities}. \label{sec-similarities}

\subsection{Simple similarities}

There is one more essential element to our schema.  It may be that
some states in $\stateset$ are equivalent to others.  That is, some states
will yield exactly the same true (or not true) interpretational statements.
Thus, we suppose that our schema comes equipped with a set
$\symgroup$ of {\em $\kinset$-similarities} (or just {\em similarities}).
Each similarity is a map $V: \stateset \rightarrow \stateset$ that
satisfies the following property.
\begin{quote}
	{\bf Property S.}  Both of these are true of $V$:
	\begin{itemize}
		\item  $V$ is a bijection.
		\item  $V D V^{-1} \in \kinset$ if and only if $D \in \kinset$.
	\end{itemize}
\end{quote}
We do {\em not} assume that every $V$ with this property
is necessarily a similarity in $\symgroup$.  However, we note
that if $V$ and $W$ satisty Property S, so does $VW$ and 
$V^{-1}$.  Thus, it is natural to suppose that 
the collection $\symgroup$ forms a group, 
and we will make that assumption.

Think of the $\kinset$-similarity map $V \in \symgroup$ as 
a set of ``spectacles'' with which we examine the states in $\stateset$.  
Through the spectacles, the state $x$ appears to be the state
$\tilde{x} = Vx$.  The dynamical law that applies the kinematically possible map 
$D$ to $x$ appears to be a different map $\tilde{D} = VDV^{-1}$, 
which is also in $\kinset$:
\begin{equation} \label{basiccommute}
    \begin{CD}
        x_{0} @>{V}>> \tilde{x}_{0}
            \\
        @V{D_{1,0}}VV  @VV{\tilde{D}_{1,0}}V
            \\
        x_{1} @>{V}>> \tilde{x}_{1}
            \\
        @V{D_{2,1}}VV  @VV{\tilde{D}_{2,1}}V
            \\
        \vdots  & &  \vdots
            \\
        @V{D_{N,N-1}}VV  @VV{\tilde{D}_{N,N-1}}V
            \\
        x_{N} @>{V}>> 
            \tilde{x}_{N}
            \\       
    \end{CD}
\end{equation}
The point is that $( \tilde{x}_0, \tilde{\vec{D}} )$ is an instance of our theory
if and only if $( x_{0}, \vec{D} )$ is.  The situation viewed through 
the spectacles fits the schema just as well as the situation without.
The spectacles simply provide a new ``frame of reference'' 
for describing the state and the time evolution.

If the theory is reversible, so that every $E \in \kinset$ has an inverse
map $E^{-1}$, we note that every element $E \in \kinset$ 
automatically satisfies Property S:
$E$ is a bijection, and $EDE^{-1} \in \kinset$ if and only if $D \in \kinset$.
This opens the possibility that the $\kinset$-similarity group $\symgroup$ might 
contain (among other things) every map in $\kinset$.  
If $\kinset \subseteq \symgroup$, we say that the the $\kinset$-similarity 
group $\symgroup$ is {\em $\kinset$-inclusive}.

A $\kinset$-similarity is not at all the same thing as a dynamical 
symmetry of a particular instance of the theory.  If $D$ is a particular
dynamical map, a dynamical symmetry $V$ would satisfy $VD = DV$, which in turn
implies that $VDV^{-1} = D$.  Property S instead has a weaker condition, that
$\tilde{D} = VDV^{-1}$ is some map in $\kinset$; but this condition must hold for every 
map $D \in \kinset$.  From a slightly different point of view, the similarity map $V$ 
acts a symmetry of the {\em sets} $\stateset$ and $\kinset$, in that
$V \stateset = \stateset$ and $V \kinset V^{-1} = \kinset$.  

Interpretational statements must respect similarities within the schema.  
For instance, suppose $P(x)$ is a state proposition in $\statementset$.  Then
for any $V \in \symgroup$, we must have $P(x) \Leftrightarrow P(Vx)$ (by which
we mean that $P(x)$ and $P(Vx)$ are true for exactly the same states $x \in \stateset$).
For a more general type of proposition,
\begin{equation} \label{eq-similarinterp}
	P(x_0,\vec{D}) \Leftrightarrow P(\tilde{x}_0,\tilde{\vec{D}}) 
		= P(Vx_0, (VD_{1,0}V^{-1}, \ldots, VD_{N,N-1}V^{-1})\,)
\end{equation}
for all $V \in \symgroup$.
Each similarity $V \in \symgroup$ imposes a restriction on the possible 
interpretational statements in $\statementset$.  Therefore, we can regard 
$\statementset$ and $\symgroup$ as ``dual'' to one another.  The larger
the set of $\kinset$-similarities, the more restricted is the allowed set of
interpretational statements.

\subsection{Extended similarities}

The similarities $V \in \symgroup$ are spectacles with which we may view
an instance of our theory.  But it is also possible to imagine time-dependent
spectacles which apply different maps at different times.  
This is analogous to translating from {\em blue-green} color
language to {\em grue-bleen} language.

What kind of time-dependent spectacles might we have?
An {\em extended similarity map} is a sequence 
$\vec{V} = (V_0, V_1, \ldots , V_{N})$ of
maps on $\stateset$.  We require that this sequence satisfies the
following property:
\begin{quote}
	{\bf Property S$\sys{ext}$.}  Both of these are true of all maps in $\vec{V}$:
	\begin{itemize}
		\item  $V_{k} \in \symgroup$.
		\item  $V_{k+1} D V_{k}^{-1} \in \kinset$ if and only if $D \in \kinset$.
	\end{itemize}
\end{quote}
The meaning of this property can be explained by a diagram.
\begin{equation}  \label{extendedcommute}
    \begin{CD}
        x_{0} @>{V_0}>> \tilde{x}_{0}
            \\
        @V{D_{1,0}}VV  @VV{\tilde{D}_{1,0}}V
            \\
        x_{1} @>{V_1}>> \tilde{x}_{1}
            \\
        @V{D_{2,1}}VV  @VV{\tilde{D}_{2,1}}V
            \\
        \vdots  & &  \vdots
            \\
        @V{D_{N,N-1}}VV  @VV{\tilde{D}_{N,N-1}}V
            \\
        x_{N} @>{V_N}>> 
            \tilde{x}_{N}
            \\       
    \end{CD}
\end{equation}
Property S$\sys{ext}$ therefore requires that, 
for an extended similarity $\vec{V}$,
$(\tilde{x}_0, \tilde{\vec{D}})$ is an instance of the theory if and only if
$(x_0,\vec{D})$ is.  We may regard $\vec{V}$ as 
a symmetry of the sets $\stateset$ and $\kinset^{N}$, 
in the sense that $V_{k} \stateset = \stateset$ and 
$V_{k+1} \kinset V_{k}^{-1} = \kinset$ for all $k$.

We denote the set of extended similarities by $\symgroup\sys{ext}$.
We do not assume that every extended map $\vec{V}$
satisfying Property S$\sys{ext}$ must be in $\symgroup\sys{ext}$.  
It is interesting to note that in some schemata 
there are examples in which $V_{k} \in \symgroup$ for all $k$, but
$\vec{V}$ fails to satisfy Property S$\sys{ext}$.
However, if $V$ satisfies Property S, then $(V, V, \ldots , V)$ 
must also satisfy Property S$\sys{ext}$.  Therefore we will assume
$(V, V, \ldots , V) \in \symgroup\sys{ext}$ for
every $V \in \symgroup$.  That is, time-independent spectacles are
always allowed in $\symgroup\sys{ext}$, and in this sense we may
say that $\symgroup \subseteq \symgroup\sys{ext}$.
We further assume that the set $\symgroup\sys{ext}$ of extended
similarities is itself a group.

An element $\vec{V}$ in the extended similarity group 
$\symgroup\sys{ext}$ turns one instance $(x_0,\vec{D})$ of a theory into another 
instance $(\tilde{x}_0,\tilde{\vec{D}})$ of the theory.  But in a more fundamental sense, 
we should regard $(x_0,\vec{D})$ and $(\tilde{x}_0,\tilde{\vec{D}})$ 
merely as different {\em pictures} 
of the same actual situation, the one picture transformed into the other by
the use of (possibly time-dependent) spectacles.
Of course, the truth of an interpretational statement should not depend on
the picture used to describe the instance of the theory.  Thus, we require that
\begin{equation} \label{eq-extsimilarinterp}
		P(x_0,\vec{D}) \Leftrightarrow P(\tilde{x}_0,\tilde{\vec{D}}) 
		= P(V_{0} x_0, (V_{1}D_{1,0}V_{0}^{-1}, \ldots, V_{N} D_{N,N-1}V_{N-1}^{-1})\,)
\end{equation}
for each $P \in \statementset$ and $\vec{V} \in \symgroup\sys{ext}$.  
We recognize this as just the extended version of Equation~\ref{eq-similarinterp}, 
and we note that it includes that fact as a special case.

We note that any extended similarity $\vec{V}$ preserves the composition
relations among the maps in $\kinset^{N}$.  Suppose for simplicity that 
our theory is reversible and we specify a particular sequence
of evolution maps $\vec{D} = (D_{1,0}, D_{2,1}, \ldots, D_{N,N-1})$.  We
define the maps $D_{kj}$ according to Equation~\ref{eq-dkjdef} and
say that  $\tilde{D}_{kj} = V_{k} D_{kj} V^{-1}_{j}$.  Then the transformed 
set of maps satisfies a transformed version of 
Equation~\ref{eq-composition}, namely that
\begin{equation}
	\tilde{D}_{k,j} = \tilde{D}_{k,l} \tilde{D}_{l,j} 
\end{equation}
for any $j,k,l \in \{0, \ldots, N\}$.  In other words, $\vec{V}$ preserves
the algebraic structure of $\kinset$ that arises from time evolution over
successive time intervals.

\subsection{The DeWitt Principle}

Our framework tells us that an interpretational system involves, not simply
the set $\statementset$ of interpretational statements, but also the 
group $\symgroup\sys{ext}$.  The former includes everything that might
be truthfully asserted about a physical situation.  The latter tells us which
different instances $(x_0, \vec{D})$ and $(\tilde{x}_{0},\tilde{\vec{D}})$ of
a theory should be regarded as different pictures of the same situation.
These are related, since the same interpretational statements must be
true in both equivalent pictures.

DeWitt's maxim says that the interpretation of quantum theory can be 
derived from the mathematical structure of the theory.  For this to hold,
we must be able to derive $\statementset$ and $\symgroup\sys{ext}$
from the mathematical structure of $\stateset$ and $\kinset$.  No 
outside elements or special assumptions need be, or should be, 
introduced.

Therefore, {\em every} map $V$ that satisfies Property S is a symmetry 
of $\stateset$ and $\kinset$, and so should be included in $\symgroup$; 
and the same is true of every sequence $\vec{V}$ of such maps satisfying 
Property S$\sys{ext}$.
Thus, we pose the following {\bf principle of maximal similarity}, which we
may for convenience call the ``DeWitt Principle''.
\begin{quote}
	{\bf DeWitt Principle.}  For a given $\stateset$ and $\kinset$, we must choose the
	similarity group $\symgroup$ and the extended group $\symgroup\sys{ext}$ to be
	maximal.
\end{quote}
That is, 
\begin{itemize}
	\item  The similarity group $\symgroup$ contains every map $V$
		satisfying Property S.
	\item  The extended similarity group $\symgroup\sys{ext}$ contains every 
		sequence $\vec{V}$ of elements of $\symgroup$ satisfying 
		Property S$\sys{ext}$.
\end{itemize}
It is not hard to show that the maximal $\symgroup$ and $\symgroup\sys{ext}$, 
as defined, exist and are groups.

When we assume that $\symgroup$ and $\symgroup\sys{ext}$ are maximal,
we maximally constrain the set $\statementset$ of interpretational statements.
This is the other side of the DeWitt Principle.  If the mathematical formalism of a theory
is capable of yielding its own interpretation, it follows that the only allowable interpretational 
statements are those that can be derived from the mathematical formalism alone.  
These interpretational statements must ``look the same'' through both time-independent
and time-dependent similarity spectacles.

Of course, as we will see, it may be that the appropriate choice
of $\symgroup\sys{ext}$ is not maximal.  There may be
additional constraints on similarities, allowing for a wider range of 
interpretational statements.  But a non-maximal choice of $\symgroup\sys{ext}$
cannot be derived from the structure of the sets $\stateset$ and $\kinset$.

\subsection{Reversibility, transitivity and interpretation}

Suppose we have a reversible theory, so that $\kinset$ is a group.  Then the DeWitt
Principle implies that every element of $\kinset$ is also a $\kinset$-similarity in $\symgroup$.
Thus, $\symgroup$ is $\kinset$-inclusive (i.e., $\kinset \subseteq \symgroup$).  And
in fact, we can say more.  In a reversible theory, 
for any sequence $\vec{E} = (E_0, E_1, \ldots , E_{N}) \in \kinset^{N}$ 
must be in $\symgroup\sys{ext}$.  Thus, $\kinset^{N} \subseteq \symgroup\sys{ext}$.

We say that the set $\kinset$ of kinematically possible maps acts
{\em transitively} on the state set $\stateset$ if, for any $x,y \in \stateset$
there exists $D \in \kinset$ so that $y = Dx$.  That is, any given
state $x$ can be turned into any other given state $y$ by some 
kinematically possible dynamical evolution.

Consider a reversible theory schema in which
$\kinset$ acts transitively on $\stateset$.  As we have seen, the 
DeWitt Principle implies that $\kinset \subseteq \symgroup$.
Any such $\kinset$-inclusive similarity group $\symgroup$ must also act
transitively on $\stateset$.  But this has an important and baleful
implication for the collection $\statementset$
of interpretational statements.  Suppose $P$ is a state proposition,
and consider two arbitrary states $x,y \in \stateset$.  
By transitivity there exists $V \in \symgroup$ such that $y = Vx$.
Thus $P(x) \Leftrightarrow P(Vx) = P(y)$.  In other words, the only possible state
propositions in $\statementset$ are those that are true for every state or for none.
There are no non-trivial state propositions in $\statementset$.

The implications for the extended similarity group $\symgroup\sys{ext}$ are even
stronger.  The DeWitt Principle applied to $\symgroup\sys{ext}$ implies that
$\kinset^{N} \subseteq \symgroup\sys{ext}$.  This means we can 
freely choose $\vec{V} \in \kinset^{N}$ and guarantee that 
$\vec{V} \in \symgroup\sys{ext}$.

Now choose any two states $x_{0},y_{0} \in \stateset$ and any two sequences
$\vec{D},\vec{E} \in \kinset^{N}$.  
Since $\kinset$ acts transitively on $\stateset$, we can find $V_{0} \in \kinset$ such that 
$y_{0} = V_{0}x_{0}$.  Furthermore, for $k \geq 1$ the map 
$V_{k} = E_{k,k-1} V_{k-1}D_{k-1,k} \in \kinset$, and so the
sequence $\vec{V}$ forms ``time-dependent spectacles'' 
in $\symgroup\sys{ext}$.  The following diagram commutes:
\begin{equation}  \label{eq-anyintoany}
    \begin{CD}
        x_{0} @>{V_0}>> y_{0}
            \\
        @V{D_{1,0}}VV  @VVE_{1,0}V
            \\
        x_{1} @>{V_1}>> y_{1}
            \\
        @V{D_{2,1}}VV  @VVE_{2,1}V
            \\
        \vdots  & &  \vdots
            \\
        @V{D_{N,N-1}}VV  @VVE_{N,N-1}V
            \\
        x_{N} @>{V_N}>> 
            y_{N}
            \\       
    \end{CD}
\end{equation}
{\em Any specific instance $(x_0,\vec{D})$ of our theory can be transformed 
into any other specific instance $(y_0,\vec{E})$.}   
Therefore, the general interpretational statements 
$P(x_0,\vec{D})$ and $P(y_0,\vec{E})$ must both be equivalent.  
This may be stated as our main general result:
\begin{quote}
	{\bf Theorem.}  Consider a reversible theory schema in which $\kinset$ 
	acts transitively on $\stateset$.  If the DeWitt Principle holds, then 
	$\statementset$ contains no non-trivial statements.
\end{quote}
We might restate this conclusion in another way:
A reversible theory in which any state could in principle
evolve to any other state {\em cannot} yield its own non-trivial
interpretation without additional constraints on $\symgroup\sys{ext}$.

\section{Examples}. \label{sec-examples}

In this section we will set up a few examples of theory schemata and discuss
some of the properties of each.  For simplicity, each example considers time
evolution over a single interval of time from $t_{0}$ to $t_{1}$.

\subsection{Deck shuffling}

Consider a standard deck of 52 cards.  The state set $\stateset$ consists of every
arrangement of the cards in the deck, and a kinematically possible map is simply a
permutation of the deck.  All such permutations are in $\kinset$.

Suppose now we divide the deck into two half-decks of 26 cards each.  Every 
rearrangement of the whole deck is in $\stateset$.  However, our kinematically
possible maps include only separate rearrangements of the half-decks.  Thus,
if the queen of hearts starts out in half-deck \#1, it will stay there no matter what
``time evolution'' $D \in \kinset$ occurs.  This, like the full-deck theory, is a 
reversible theory.

The DeWitt Principle implies that $\kinset \subseteq \symgroup$ for both theories.
For the undivided deck, the permutation group acts transitively on the state set.
This theory, therefore, has no non-trivial statements in $\statementset$.

What about $\symgroup$ and $\symgroup\sys{ext}$ for the half-deck theory?
In this schema there are maps in the maximal $\symgroup$ that are
not in $\kinset$.  For instance, consider a map $X$ on states that exchanges the two
half-decks.  This is not in $\kinset$, but it does satisfy Property S since 
both $XDX^{-1}$ and $X^{-1}DX$ are half-deck shuffles.  (The two half-decks 
are exchanged twice.)  From the DeWitt Principle, both $X$ and the 
identity map $1$ are in $\symgroup$.  However, the sequence $\vec{V} = (1,X)$ does
not satisfy Property S$\sys{ext}$ and therefore is not in $\symgroup\sys{ext}$.

In the half-deck theory, $\kinset$ is a group but it does not act transitively
on $\stateset$.  The divided deck with separate half-deck permutations 
does potentially have non-trivial statements in $\statementset$.  For example, 
the statement ``All of the jacks are in the same half-deck'' will not change 
its truth value if the half-decks are reshuffled or exchanged.  Such a statement
expresses a property that may be the basis for an interpretational statement.

\subsection{Symbolic dynamics}

A very interesting example arises from {\em symbolic dynamics}.  In symbolic
dynamics, the states are bi-infinite sequences of symbols from a finite alphabet.
The set of allowed sequences may be constrained by some rule; for instance, we may
be restricted to binary sequences that never include more than two 1's in succession.
The particular
example we will consider includes all binary sequences in $\stateset$.  This is known
in the literature as the ``full shift'' and is
the symbolic dynamics associated with the ``baker's map'' on the unit square.

The dynamical maps are finite left or right shifts of the sequences in $\stateset$.  
There are thus two reasonable choices for $\kinset$.  
First, $\kinset$ might contain only the elementary map $\sigma$ 
that shifts the sequence by one place:  given a sequence $\vec{x}$,
$(\sigma \vec{x})_{i} = x_{i+1}$.  
Second, we might posit that $\kinset$ includes all finite shifts, so that
$\kinset = \{ \ldots, \sigma^{-1}, 1, \sigma, \sigma^{2}, \ldots \}$.
This amounts to assuming that the underlying time evolution can occur at 
any finite speed, so that an arbitrary number of elementary shifts 
in either direction may occur within our given time interval.

We will make the second choice, which makes $\kinset$ a group and the theory reversible.
Thus, under the DeWitt Principle, all the shifts in $\kinset$ are also similarities
in the maximal group $\symgroup$.  This maximal $\symgroup$ also includes many 
other maps as well.  For example, $\symgroup$ contains the map $\beta$ 
that complements the sequence:  
$(\beta \vec{x})_{i} = \bar{x}_{i}$, where $\bar{0} = 1$ and $\bar{1} = 0$.
It also contains the reflection map $\rho$:  $(\rho \vec{x})_{i} = x_{-i}$. 
However, $\symgroup$ cannot contain any map $V$ that takes 
a constant sequence to a non-constant sequence.

Let us prove this assertion.  Our definition of the similarity group
$\symgroup$ for symbolic dynamics implies the following:
\begin{quotation}
	\noindent
	If $V \in \symgroup$, then for all $n \in \mathbb{Z}$ there exists
	$m \in \mathbb{Z}$ such that $V^{-1} \sigma^{n} V = \sigma^{m}$,
	or equivalently $\sigma^{n} V = V \sigma^{m}$.
\end{quotation}
We will use the contrapositive of this fact.
\begin{quotation}
	\noindent
	If there exists $n \in \mathbb{Z}$ such that for all $m \in \mathbb{Z}$
	we have $\sigma^{n} V \neq V \sigma^{m}$, then $V \notin \symgroup$.
\end{quotation}
Now consider the constant sequence $\vec{b} = \ldots bbbb \ldots$, and
suppose $V \vec{b}$ is not constant.  Then there exists $n \in \mathbb{Z}$
such that $\sigma^{n} V \vec{b} \neq V \vec{b}$.  
But for any $m \in \mathbb{Z}$, $\vec{b} = \sigma^{m} \vec{b}$, and
so $\sigma^{n} V \vec{b} \neq V \sigma^{m} \vec{b}$.  Thus 
$\sigma^{n} V \neq V \sigma^{m}$, and hence $V \notin \symgroup$.

The similarity group $\symgroup$ does not act transitively on $\stateset$.
Therefore, even if we impose the DeWitt Principle, 
the statements in $\statementset$ may still include nontrivial statements 
like, ``The sequence is constant,'' which retain their truth value under
shifts, reflection, complementation, etc.

\subsection{Classical Hamiltonian dynamics}

Suppose we have a classical system described by a phase space with $n$ real coordinates
$q_{k}$ and $n$ associated momenta $p_{k}$.  To make things a bit simpler, we can
shift our time coordinate so that $t_{0} = 0$ and $t_{1} = \tau$.
The allowed time evolutions in $\kinset$ are 
the ``Hamiltonian maps'' that result from a (possibily time-dependent) Hamiltonian function 
$H(q_{k}, p_{k}, t)$ acting over the time interval ($t=0$ to $t=\tau$), so that
\begin{equation}
	\dot{p}_{k} = \frac{dp_{k}}{dt} = - \frac{\partial H}{\partial q_{k}} \qquad \mbox{and} \qquad
	\dot{q}_{k} = \frac{dq_{k}}{dt} = \frac{\partial H}{\partial p_{k}}  .
\end{equation}
Two maps can be composed as follows.  Suppose we have maps $D_{1}$ and $D_{2}$, which
are produced by Hamiltonian functions $H_{1}(q_{k},p_{k},t)$ and $H_{2}(q_{k},p_{k},t)$
controlling the dynamics over the time interval $0$ to $\tau$.  
Then we can construct a new map $D_{21}$ 
via the following Hamiltonian:
\begin{equation}
	H_{21}(q_{k},p_{k},t) = \left \{  
		\begin{array}{ll}
		2H_{1}(q_{k},p_{k},2t)  & 0 \leq t \leq \tau/2 \\
		2H_{2}(q_{k},p_{k},2t-\tau)  & \tau/2 < t \leq \tau .
		\end{array} \right .
\end{equation}
This will cause the system to evolve according to a ``two times faster'' version
of $H_{1}$ for the first half of the time interval, and a ``two times faster'' version
of $H_{2}$ for the second half of the interval. The resulting change in state will
simply be the map $D_{21} = D_{2} D_{1}$.

This theory is reversible, since the evolution by $H(q_{k}, p_{k}, t)$ can be 
exactly reversed by the Hamiltonian $-H(q_{k}, p_{k}, \tau-t)$.  Thus
the maximal $\symgroup$ includes all of $\kinset$, and potentially many
other maps.

The set of Hamiltonian maps also acts transitively on the classical phase space.  
Given any two points $(q_{k},p_{k})$ and 
$(q_{k}',p_{k}')$, it is not hard to write down a Hamiltonian
function that evolves one into the other in the time interval 
from $0$ to $\tau$.  Thus, if the DeWitt Principle holds,
$\statementset$ contains no non-trivial interpretational statements.

\subsection{Unitary quantum mechanics}

In quantum theory, the states in $\stateset$ are vectors $\ket{\psi}$ of unit norm 
in a Hilbert space $\hilbert$.  As before, we take $\dim \hilbert$ to be finite, 
though maybe extremely large.  The kinematically possible maps $\kinset$ include
all unitary operators on $\hilbert$.  All such operators can be realized by
evolving the state vector via the Schr\"{o}dinger equation using
the Hamiltonian operator $\oper{H}(t)$:
\begin{equation}
	i \hbar \ket{\psi(t)} = \oper{H}(t) \ket{\psi(t)}  \qquad \Longrightarrow \qquad
	\ket{\psi(t_{1})} = \oper{U} \ket{\psi(t_{0})} .
\end{equation}
Since this theory is reversible, the maximal similarity group $\symgroup$
includes all of the unitary operators in $\kinset$.
The unitary operators also act transitively
on the unit vectors in a Hilbert space $\hilbert$.  Thus, the DeWitt
Principle excludes all non-trivial interpretational statements
from $\statementset$.

From these examples we may draw a general lesson.
Some theories have non-trivial statements whose truth value
is unchanged by any similarity, even when $\symgroup$ and
$\symgroup\sys{ext}$ are maximal.  In this way, it is possible
that ``the mathematical formalism" of a theory could yield
``its own interpretation''.  
But this is impossible for many interesting
theories, including both classical Hamiltonian
dynamics and unitary quantum mechanics.

\section{Taming quantum similarities?}. \label{sec-taming}

Suppose we have a reversible theory schema in which $\kinset$ acts
transitively on $\stateset$.  Under the DeWitt Principle, the 
unlimited similarity groups $\symgroup$ and $\symgroup\sys{ext}$ 
are too big to admit non-trivial interpretational statements in
$\statementset$.  Therefore, any meaningful interpretation
for the theory will require us to limit the similarity groups in
some way.  We must either have $\kinset \not\subseteq \symgroup$
or $\kinset^{N} \not\subseteq \symgroup\sys{ext}$, or both.
This is precisely the ``additional structure'' posited by Wallace \cite{wallace},
discussed in Subsection~\ref{subsec-whatisasystem} above.

The basis for a limitation of this kind cannot be found in the 
mathematical formalism of $\stateset$ and $\kinset$.  Any such
external limitation will therefore contravene our version of the 
DeWitt Principle.  It will be useful here briefly to describe 
a couple of plausible ``non-DeWitt'' limitations on 
$\symgroup$ and $\symgroup\sys{ext}$
for the example of unitary quantum mechanics over a 
single time interval.  

\subsection{Subsystem decomposition}
First, suppose $\hilbert$ can be decomposed
as a tensor product of smaller spaces:
$\hilbert = \hilbert\sys{1} \otimes \hilbert\sys{2} \otimes \cdots \otimes \hilbert\sys{n}$.
(This is one of the possibilities mentioned by Wallace.)
Each $\hilbert\sys{k}$ represents the state space of a subsystem
of the whole quantum system.  This does not by itself limit the kinematically
possible time evolutions in $\kinset$, since the subsystems might interact
with one another in an arbitrary way.  But if we take the subsystem
decomposition as given, we may plausibly restrict our similarities
to operators of the form:
\begin{equation}  \label{subsystemsimilarity}
	V = V\sys{1} \otimes V\sys{2} \otimes \cdots \otimes V\sys{n} .
\end{equation}
Our similarity spectacles can modify the states of the individual
subsystems, but they cannot mix the subsystems together.  In this
case, even though $\kinset$ acts transitively on $\stateset$, 
the similarity group $\symgroup$ does not.  This restriction on
$\symgroup$ (and hence $\symgroup\sys{ext}$) allows for 
many non-trivial interpretational statements in $\statementset$.
For example, consider the state proposition 
$P(x)$ = ``In state $x$, subsystems 1 and 2 are entangled.''  
Since the $\kinset$-similarities do not mix subsystems, 
this statement has the same truth value, 
regardless of what similarity spectacles are applied to the state.

We must remember, however, that there are infinitely many tensor product 
decompositions of $\hilbert$ \cite{meronomic}.  
That is, we can decompose a composite 
system into subsystems in an unlimited
number of ways.  States that are entangled with respect to
one decomposition may not be entangled with respect to 
another.  For instance, consider a system with $\dim \hilbert = 4$
that can be regarded as a pair of qubits, labeled 1 and 2.
This pair could be in one of the four entangled ``Bell states'':
\begin{equation}
	\begin{array}{l}
	\ket{\Phi_{\pm}\sys{12}} = \frac{1}{\sqrt{2}} 
		\left ( \ket{0\sys{1}} \otimes \ket{0\sys{2}} \pm
		\ket{1\sys{1}} \otimes \ket{1\sys{2}} \right )
	\\[1ex]
	\ket{\Psi_{\pm}\sys{12}} = \frac{1}{\sqrt{2}} 
		\left ( \ket{0\sys{1}} \otimes \ket{1\sys{2}} \pm
		\ket{1\sys{1}} \otimes \ket{0\sys{2}} \right )	 .
	\end{array}
\end{equation} 
On the other hand, there exists an entirely different decomposition
of the system into qubits designated A and B, with respect to which
these are product states:
\begin{equation}
	\begin{array}{lcl}
	\ket{\Phi_{+}\sys{12}} = \ket{\Phi\sys{A}} \otimes \ket{+\sys{B}}
	& \quad & 
	\ket{\Psi_{+}\sys{12}} = \ket{\Psi\sys{A}} \otimes \ket{+\sys{B}}
	\\
	\ket{\Phi_{-}\sys{12}} = \ket{\Phi\sys{A}} \otimes \ket{-\sys{B}}
	& \quad & 
	\ket{\Psi_{-}\sys{12}} = \ket{\Psi\sys{A}} \otimes \ket{-\sys{B}} .
	\end{array}
\end{equation}
Subsystem decompositions are necessary to describe many 
important processes.  For example, decoherence processes
depend on the decomposition of the whole system into a
subsystem of interest and an external environment.

We must therefore ask, where does a special subsystem
decomposition come from?
Neither the set of possible states $\stateset$ nor the set $\kinset$ 
of kinematically possible maps picks out a particular
decomposition.  It must come from somewhere else.  
Non-trivial interpretational statements about entanglement
are only possible once a preferred decomposition is specified,
by whatever means.

From the point of view espoused by Wallace \cite{wallace}, 
the subsystem decomposition is simply a {\em given} for a particular 
physical situation.
The mathematical formalism of quantum theory specifies $\stateset$
and $\kinset$ {\em and} a similarity group $\symgroup$ that
respects the preferred subsystem decomposition.  
The question of the physical basis for this
decomposition---its origin and representation in the state 
and dynamics of the system of interest---simply cannot arise.
As Wallace himself points out, however, this decomposition 
is itself the real source of the complexity of the quantum
world.

If we allow ourselves to invoke a hypothetical outside observer, it is
easy to see how a preferred decomposition could emerge.  
The subsystems in the special decomposition correspond 
to different ways that the observer can access the system of interest.  
{\em This} sort of control or measurement interaction affects {\em this} 
subsystem, {\em that} sort affects {\em that} subsystem.  The
decomposition emerges from the nature of the devices that
implement these operations.  But these devices do {\em not} 
reside in the system of interest, and their intervention means
that the system is no longer isolated.

Subsystem decomposition is a special type of quantum reference
frame information, called {\em meronomic} information \cite{meronomic}.  We 
will briefly discuss the role of quantum reference frames in 
Subsection~\ref{ssec-qrfs} below.

\subsection{Time-independent spectacles}

Here is another potential limitation, this one on the extended
similarity group $\symgroup\sys{ext}$.  We allow any unitary
map $V \in \symgroup$, but we declare that the only elements
of $\symgroup\sys{ext}$ are those of the form $(V, V)$.
Only ``time-independent spectacles'' are allowed; no ``grue-bleen''
pictures are permitted.  In this case,
$\symgroup$ acts transitively on $\stateset$, and only trivial
state propositions $P(x)$ are possible in $\statementset$.
However, there are non-trivial general propositions in $\statementset$.
For example, consider the statement $Q(x,D) = $ ``State $x$ is a fixed
point of dynamics $D$; that is, $Dx = x$.''  If we apply the (time-independent)
similarity map $V$ to turn instance $(x,D)$ into $(\tilde{x}, \tilde{D})$, 
we find that $\tilde{D} \tilde{x} = VDV^{-1} V x = V D x = V x = \tilde{x}$.
The statement $Q(x,D)$ might be true or not---it is not trivial---but
in any case $Q(x,D) \Leftrightarrow Q(\tilde{x}, \tilde{D})$.

Even for a schema with a single time interval, we are effectively
dealing with {\em two} sets of states:  $\stateset_{0}$ at $t_{0}$ and
$\stateset_{1}$ at $t_{1}$.  These are of course both isomorphic to
$\stateset$.  One connection between the sets is the dynamical 
evolution $D \in \kinset$, which indicates which $x_{0} \in \stateset_{0}$
evolves to $x_{1} \in \stateset_{1}$.  To claim that our spectacles
are ``time-independent'' means that we have another canonical 
isomorphism between the two, which lets us identify which states
in $\stateset_{0}$ are taken to be {\em identical} to other states 
in $\stateset_{1}$.  We might denote this canonical isomorphism
by the symbol 1, but this hides the fact that there are {\em infinitely 
many} possible isomorphisms between the two sets.  
To say unambiguously that a state at $t_{0}$ is the
same state as another at $t_{1}$, or to define some spectacles
as ``time-independent'', we must invoke this second way (besides the
time evolution map $D \in \kinset$) to link together $\stateset_{0}$
and $\stateset_{1}$.

We might, of course, simply argue that this link between $\stateset_{0}$
and $\stateset_{1}$ is part of the {\em definition} of the system
of interest.  But if we do not regard this answer-by-definition as
satisfactory, the question remains:
What is the physical origin of such a link, which is required 
to make the needed restrictions on $\symgroup\sys{ext}$?  
If the quantum system is truly isolated, no satisfactory answer is possible,
since $D$ itself describes how all parts of the state evolve, and thus
expresses everything about the dynamical connection between
times $t_{0}$ and $t_{1}$.
But once again, a hypothetical outside observer can 
provide a plausible answer.  The
external apparatus of the observer can allow us to define
what it means for a state to remain the same over time.  
In effect it provides a fixed reference frame for the Hilbert 
space of states.

Such an explanation seems natural, but of course it invokes an
observer that is {\em not} treated as part of the isolated
quantum-mechanical system.  It runs counter to the letter
and spirit of DeWitt's maxim.

\subsection{Quantum reference frames} \label{ssec-qrfs}

Ours is essentially a reference frame problem, so it is natural
to ask whether the existing theory of quantum reference frames
\cite{qref} can help resolve it.  Unfortunately, it cannot.

In quantum reference frames, we begin with an abstract 
symmetry group $\mathcal{G}$.  
Any system is made of up of elementary subsystems,
each of which has its own unitary representation of $\mathcal{G}$.
The symmetry element $g \in \mathcal{G}$ is represented by
the unitary operator
\begin{equation}
	\oper{V}_{g} = \oper{V}\sys{1}_{g} \otimes \oper{V}\sys{2}_{g}
		\otimes \cdots \otimes \oper{V}\sys{N}_{g} 
\end{equation}
for subsystems 1, \ldots, N.
These operators are dynamical symmetries for the system, 
so that the only available operations are symmetric ones, 
those that commute with $\oper{V}_{g}$.  
Nevertheless, if part of the system is
in an asymmetric state, we can use that state as a
resource to perform asymmetric operations on other 
parts of the system.  
This asymmetric resource state constitutes a
quantum reference frame.

To take an example, suppose our subystems are spin-1/2 particles
and our symmetry group $\mathcal{G}$ is the set of rotations
in 3-D space.  Each spin has its own $SU(2)$ representation of
this group.  We can only perform rotationally invariant operations
on the spins.  A measurement of $\oper{S}\sys{1}_{z}$ on 
spin \#1 thus seems out of the question, since we cannot {\em a priori}
specify the $z$-axis.  However, suppose the remaining N-1
spins are provided in the state $\ket{\uparrow\sys{k}}$, aligned with 
the (unknown) $z$-axis.  Then we can use these extra spins
to perform a global rotationally invariant operation that approximates
an $\oper{S}\sys{1}_{z}$ measurement on the first spin.  
We have used the asymmetric $\ket{\uparrow\sys{k}}$ states as 
a quantum reference frame resource.

The decomposition of a quantum system into subsystems can also
be described as a quantum reference frame problem \cite{meronomic}.  
For example, suppose we consider some quantum systems
with $\dim \hilbert = 4$ (called ``tictacs'' in \cite{meronomic}), and we wish
to specify a particular subsystem decomposition for these into
qubit pairs.  We can do this by supplying additional tictacs in a 
special ``asymmetric'' state that encodes the
subsystem division.  For example, suppose we are considering
a series of tictacs in state $\ket{\Phi}$, and we wish to 
estimate the Schmidt parameter of the entangled state for a
particular qubit decomposition.  We can accomplish this with
the assistance of a supply of tictac pairs in the resource state
$\ket{\Psi_{-}\sys{13}} \otimes \ket{\Psi_{-}\sys{24}}$ (where
the first tictac is made up of qubits \#1 and \#2 and the second
is made up of \#3 and \#4).  If we specify how to decompose a
particular system into subsystems, we say that we have provided
{\em meronomic} frame information.  We therefore see that
meronomic information for dividing tictacs into qubits 
can be regarded as a kind of quantum information, 
information that can in principle be represented by the state of
quantum systems.

The symmetry group $\mathcal{G}$ (or more precisely its unitary
representation $\{ \oper{V}_{g} \}$) is somewhat analogous 
to our similarity group $\symgroup$.  While the symmetry element
$g$ remains unknown, we can only make $\mathcal{G}$-invariant
statements about our system.  Notice that if we add new
subsystems to our system, we do not actually enlarge the symmetry
group.  The symmetry group for N spins is still just a representation 
of $SU(2)$.  Informally, we may say that the ``symmetry frame problem'' 
stays essentially the same when we enlarge the system,
but the additional pieces may provide asymmetric states
as resources to help resolve the problem.  

However, under the DeWitt Principle, the similarity group 
$\symgroup$ for N spins contains all of $U(2^{\mbox{\tiny N}})$,
the full set of unitary operators on the Hilbert space for
the spins.  The ``similarity frame problem'' gets {\em worse} as we 
add spins, not better.  Even if we are somehow granted 
the subsystem decomposition between the spins,
so that the similarity group contains $U(2) \otimes U(2) \otimes
\cdots \otimes U(2)$, the state of the final N-1 spins can provide 
{\em no information} about the similarity frame of spin \#1.

This problem is already present for meronomic frame information.
We can provide quantum resources for specifying how tictacs
can be divided into qubits, but this protocol presumes that the
decomposition of the world into tictacs is already given.  That
decomposition can be encoded into states of even larger systems,
but at every stage we must presume the decomposition of a 
bigger universe into larger chunks.  The meronomic frame 
problem gets worse as we introduce more quantum resources 
to resolve it.

\section{Remarks}. \label{sec-remarks}

We have avoided giving a formal definition of the ``interpretation''
of a theory.  But informally, we might say that an interpretation is 
a set of rules for extracting meaning from the mathematical formalism 
of a theory.  In quantum mechanics, the formalism includes a 
global quantum state that evolves unitarily.  The many-worlds 
interpretation claims to extract from this formalism various 
meaningful statements about processes and correlations, 
including observations made by observer subsystems.

The problem is that any mathematical framework of states and
time evolution maps ($\stateset$ and $\kinset$) entails a 
group of automorphisms, which we have called ``similarities''.
These similarities may be time-independent, or they may be 
time-dependent (like the shift from {\em green/blue} color
language to {\em grue-bleen} color language).
When viewed through the spectacles of a similarity transformation,
one particular instance of a theory is transformed into another.
In some cases---including unitary quantum mechanics---{\em any}
instance can be transformed into {\em any} other.

The complex universe Q of Section~\ref{subsec:twouniverses}
seems very different from the simple universe Q', and any
interpretational approach that cannot distinguish them is
plainly inadequate.  Yet the two universes are related by a 
similarity transformation of the underlying theory---they are,
in effect, two pictures of the {\em same} universe.  How 
is our interpretation to distinguish them?  The only way to
fix this problem is to impose a restriction on the set of similarities.

If we regard quantum theory as a pragmatic set of rules
that an observer applies to analyze a limited, external system, then
such a restriction is reasonable.  It may arise, not from anything
``inside'' the system itself, but from the relationship
between the observer and the system.
The observer may well insist on this additional structure
before applying the theory.
But the many-worlds program requires that we regard quantum
theory as a description of an entire universe that
includes the observer.  Recall that Everett titled his detailed
account ``The Theory of the {\em Universal}
Wave Function'' (\cite{everett1957}, emphasis ours).  

We are left with a quandary.  We must appeal to additional
``frame'' information beyond $\stateset$ and $\kinset$ in order 
to apply quantum theory in a meaningful way.  
This information is not quantum information---that
is, information residing in the state of the system of interest.  
The interpretational frame is not a quantum reference frame.
But if we simply require this frame information on pragmatic grounds, 
as a mere prerequisite for applying the theory, we have forfeited one 
of the central motivations of the many-worlds interpretation.  
Inasmuch as the many-worlds program
aims to implement DeWitt's maxim---that the mathematical
formalism of quantum mechanics can yield its own 
interpretation---that program fails.

The reader may wonder whether this is simply a new type of
many-worlds situation.  Perhaps every different possible 
``picture'' of an evolving quantum system is equally meaningful, 
and a full interpretation embraces them all.  But this will not do.
The ``worlds'' represented in a quantum state correspond to
distinct branches or superposition components of the global
quantum wave function.  The different branches evolve independently
according to a given time evolution $\oper{U}(t)$.  
This allows us to make conditional predictions,
e.g., ``Given that the observer's record of the previous spin 
measurement is that $S_{z} = + \mbox{$\frac{\hbar}{2}$}$,
the next measurement will yield the same result.''  
But the many-pictures idea supports no sort of predictability at all.
All possible time-evolutions, including those with wildly
varying Hamiltonians $\oper{H}(t)$, are equally admissible pictures
of the same universe.  We cannot use the past behavior of the 
universe, or our present records of that behavior, to make
any reliable prediction of future events.  A many-pictures
approach can yield no meaningful interpretation.

We have seen some simple theories (e.g., symbolic dynamics) in which
non-trivial interpretational statements are possible even with maximal
similarity groups $\symgroup$ and $\symgroup\sys{ext}$.  On the other hand,
the same difficulties do arise in classical Hamiltonian mechanics.
This has not usually been recognized as a problem because 
the ordinary classical dynamical variables---for instance,
the relative positions of particles in space---are 
generally assumed to have immediate physical meanings.  
Only with the introduction of quantum mechanics 
are interpretational issues recognized.

Obviously, we are able to use both classical and quantum mechanics
to analyze the behavior of systems, extracting meaningful 
interpretational statements.  We resolve the similarity
problem, just as we resolve the {\em grue-bleen} color
language problem, by appealing to objects and procedures that
are not contained within the system of interest.  In this view, we always
interpret quantum mechanics by appealing, implicitly or 
explicitly, to sectors of the 
universe that are not treated as parts of the 
quantum system.  In so doing, we presume that 
these external entities do not themselves
have interpretational ambiguities.  Their dynamical variables
have immediate physical meaning; their reference frames
for subsystem decomposition and time evolution are
given.  They provide our frame for interpreting the quantum
physics of the system of interest.  And this is true even if 
we formally adopt a many-worlds view of the system and its
behavior.  Or to put the same point another way, 
{\em a truly isolated quantum system has no interpretaiton.}

In this paper we have not proposed or endorsed any particular
interpretation of quantum mechanics.  Many interpretations seem
to offer valuable insights; none of them seem entirely satisfactory.
Our point is simply that any successful interpretation---any 
interpretation that generates non-trivial interpretational statements
about a theory---must somehow limit the similarity groups 
$\symgroup$ and $\symgroup\sys{ext}$ for that theory.  However, the
mere mathematical structure of Hilbert space and unitary 
operators does not appear to offer a way to do this.  We
are fully in agreement with Wallace's cautionary remark about
``additional structure''.  Without
a resolution of the quantum ``grue-bleen'' problem, no meaningful
interpretation is possible.

The traditional ``Copenhagen'' interpretation of quantum mechanics
relies on a conceptually independent macroscopic ``classical'' domain
\cite{copenhagen}.
The interaction of subsystems becomes a measurement when the
measurement record is irreversibly amplified into this domain.
The quantum evolution of an isolated system has no meaning
except that given by the possible results of such measurement processes.
As John Wheeler said, ``No elementary phenomenon is a phenomenon
until it is an observed phenomenon.''\cite{nophenomenon}

Thus, although we do not defend any particular interpretation,
our considerations here lead us toward a Copenhagen-style 
point of view.  
In some theories, including quantum mechanics, we simply cannot 
construct a viable interpretation of a system based only on the 
states and dynamical evolution of the system itself.  The physical
basis for any interpretation must lie outside the system---not 
necessarily as a separate ``classical'' domain, but as a domain 
that is somehow excluded from the similarity transformations
implicit in the mathematical formalism of the theory.

An analogy to our situation may perhaps be found in axiomatic set theory.
Given any set $X$, a larger one can be found (e.g., by forming the 
power set $\mathcal{P}(X)$).  Thus, there is no upper limit to the size 
of the objects describable in the theory.  However,
the collection of all sets is not a self-consistent set.  The ``universe''
of set theory is not an object within the theory \cite{settheory}.  

Perhaps something similar holds for physical theories like quantum
mechanics.  There is no fundamental limit to the size of the system
that can have a non-trivial interpretation.  Even a large system could 
be embedded in a still larger system that provides the necessary 
interpretational frame.  If we in turn wish to treat the larger system
within the theory, we can (in principle) embed it in a simply enormous
``super-system'' to fix its frame.  However, it is not possible to have
a non-trivial interpretation for a quantum system that includes the entire
universe.

The authors gratefully acknowledge many helpful comments from
Chris Fuchs, Rob Spekkens, Bill Wootters, Austin Hulse, and Fred
Strauch.  They are of course not responsible for the remaining shortcomings
of this paper.


\begin{thebibliography}{99}
%
\bibitem{everett1957}  H. Everett, ``Relative State Formulation of Quantum
	Mechanics'', {\em Rev. Mod. Phys.} {\bf 29}, 454--462 (1957);
	see also ``The Theory of the Universal Wave Function'' in B. S. DeWitt and N.
	Graham, {\em The Many-Worlds Interpretation of Quantum Mechanics}
	(Princeton University Press, Princeton, 1973).
%
\bibitem{bornrule}  J. B. Hartle, ``Quantum Mechanics of Individual Systems'',
	{\em Am. J. Phys.} {\bf 36}, 704-712 (1968).  W. H. Zurek, ``Probabilities
	from Entanglement, Born's rule $p_{k} = | \psi_{k} |^{2}$ from 
	Envariance'', {\em Phys. Rev. A} {\bf 71}, 052105 (2005).
%
\bibitem{zurek}  W. H. Zurek, ``Environment-Induced Superselection Rules'',
	{\em Phys. Rev. D}  {\bf 26}, 1862--1880 (1982).
%
\bibitem{quasi}  J. P. Paz and W. H. Zurek, ``Environment-induced decoherence and
	the transition from quantum to classical'', in R. Kaiser, C. Westbrook and F. David (eds.),
	{\em Decoherence:  Theoretical, Experimental, and Conceptual Problems}
	(Springer, Berlin, 2000).  E. Joos, H. D. Zeh, C. Keifer, D. Guilini, J. Kupsch,
	and I.-O. Stamatescu, {\em Decoherence and the Appearance of a Classical
	World in Quantum Theory} (2nd ed.) (Springer, New York, 2003).
%
\bibitem{dewitt}  B. S. DeWitt, ``Quantum Mechanics and Reality'', {\em Physics
	Today} {\bf 23}, 30--35.
%
\bibitem{wheeler}  John A. Wheeler, ``Assessment of Everet's `Relative State' 
	Formulation of Quantum Theory'', {\em Rev. Mod. Phys.} {\bf 29}, 463--465.
%
\bibitem{graham}  N. Graham, ``The Measurement of Relative Frequency'', in
	B. S. DeWitt and N. Graham (eds.), {\em The Many-Worlds Interpretation
	of Quantum Mechanics} (Princeton University Press, Princeton, 1973).
%
\bibitem{asher}  Asher Peres, {\em Quantum Theory:  Concepts and Methods} (Kluwer
	Academic Publishers, Dordrecht, 1995).
%
\bibitem{goodman}  N. Goodman, {\em Fact, Fiction, and Forecast} (Harvard University
	Press, Cambridge, 1955).
%
\bibitem{wallace} D. Wallace, {\em The Emergent Multiverse:  Quantum Theory
	according to the Everett Interpretation} (Oxford University Press, Oxford, 2012).
%
\bibitem{meronomic}  A. Hulse and B. Schumacher, ``Quantum meronomic frames'',
	arXiv:quant-ph/1907.04899.
%
\bibitem{qref}  S. D. Bartlett, T. Rudolph and R. W. Spekkens, ``Reference frames,
	superselection rules, and quantum information'', {\em Rev. Mod. Phys}
	{\bf 79}, 555--609 (2007).
%
\bibitem{copenhagen}  N. Bohr, ``The quantum postulate and the recent development of 
	atomic theory'', {\em Nature} {\bf 121}, 580--590 (1928).  W. Heisenberg, ``The 
	Development of the Interpretation of the Quantum Theory'', in W. Pauli (ed.),
	{\em Niels Bohr and the Development of Physics} (Pergamon, London, 1955).
%
\bibitem{nophenomenon}  J. A. Wheeler, {\em Frontiers of Time} (North-Holland, Amsterdam,
	1979).
%
\bibitem{settheory}  Paul R. Halmos, {\em Naive Set Theory} (Van Nostrand, Princeton, 1960).
%
\end{thebibliography}
\end{document}